# Basic Aspects of Negator Algebra in SOC

Rainer E. Zimmermann

IAG Philosophische Grundlagenprobleme,
FB 1, UGH, Nora-Platiel-Str.1, D – 34127 Kassel /
Clare Hall, UK – Cambridge CB3 9AL[1] /
Lehrgebiet Philosophie, FB 13 AW, FH,
Lothstr.34, D – 80335 München[2]
e-mail: pd00108@mail.lrz-muenchen.de

**Abstract**

Recent developments in loop quantum gravity and topological field theory are being mirrored with a view to the emergent structure of self-organized criticality (SOC). Referring back to an earlier paper [1], the relationship of SOC to negator algebra is discussed. It is shown that introducing the categorial perspective leads to further holistic conclusions of considerable universality. This present paper shall serve as a preparation of a concrete research project under way designed to illustrate this very universality being reflected in a complex application which is situated far away from the actual field of physics. [2] Hence, this paper belongs to a series of recent publications discussing the modern and fruitful interaction between philosophy and the sciences, beyond mere historical aspects and the traditional rephrasing of scientific results. [3]

**Introduction**

Originally, negator algebra had been introduced in order to model the formation of structure in global terms, visualizing the whole of the Universe as a permanently recursive process. The basic idea was simple enough: A sufficiently complex dynamic, prominent from examples given by the Brussels school of Prigogine in the early eighties of the last century (and actually dating back as far as to the early seventies), has been the (biological) *Keller-Segel scenario* dealing with the life cycle of the slime mold. This dynamic can be essentially modelled in terms of two equations which are of the reaction-diffusion type introduced by the Brussels school: If $a$ is the (particle number) density of amoebae constituting the initial population subject to the spontaneous onset of structure formation, and if $\rho$ is the appropriate density of the acrasine which serves as the communi-cative mediator among the amoebae (due to chemotaxis) eventually signalling the moment of agglomeration (then triggering the onset of forming a slime mold), then the evolution equations of the dynamic can be given in the form

$$\partial a/\partial t = - \nabla (D_1 \nabla \rho) + \nabla (D_2 \nabla a),$$

$$\partial \rho/\partial t = - k(\rho)\rho + a f(\rho) + D_\rho \nabla^2 \rho.$$

---

[1] Permanent addresses.
[2] Present address.

Here k and f refer to production terms of the acrasine while the Ds display the diffusion effects. In particular, $D_1$ refers to the chemotactic communication among amoebae. As has been discussed in detail at another place [4], a perturbation analysis around the equilibria of the above equations uncovers a criterion for the onset of instability such that he dynamic becomes unstable, if

$$D_1\rho_0/D_2 a_0 + a_0 f'(\rho_0)/c > 1,$$

where the zero index refers to the respective equilibria and c is a composite constant of the form $c = k(\rho_0) + \rho_0 k'(\rho_0)$, the prime referring to derivatives with respect to $\rho$. This threshold criterion is not only indicating a self-organizing property of the underlying population (namely to eventually accumulate once the food sources in the population's vicinity have become scarce), but is also the characteristic for the actual transition of the collective population to another world state (level) which is given by the slime mold as the result of the accumulation. Hence, the population (e)merges onto a higher level of organization thus creating a new organism (structure). The mold is transporting itself (and hence the population) to other places, where it spreads spores over the environment which in turn become amoebae again beginning a new cycle. If visualized as a two-component vector system with $x = (\rho, a)$, the above equations can be thought of as a differential operator matrix E (evolution operator) acting upon x by means of the product

$$Ex = 0,$$

where the appropriate matrix components are

$$E_{11} = - \nabla (D_1 \nabla), \quad E_{12} = \nabla (D_2 \nabla) - \partial/\partial t,$$

$$E_{21} = -k(\rho) + D_\rho \nabla^2 - \partial/\partial t, \qquad E_{22} = f(\rho).$$

If E is this evolution operator for a stable configuration of coefficients, and E* is the same for an unstable configuration satisfying the above criterion, then we can think of E* as that configuration of states which negates stability. Hence, the name *negator* ( = negation operator). The important point however is the following: The transition E → E* represents nothing but the onset of instability, but not yet the new (stable) structure, because this is being described on a higher level of evolution, i.e. by a different dynamic. Assume that this new dynamic can be modelled by another evolution operator of the form E**. Then it is really the full transition sequence E → E* → E** which gives the actual formation of structure (or the respective phenomenon of emergence). If we represent the negation formally by E* = N(E), then E** = N$^2$(E), i.e. the „negation of the negation." (The terminology is chosen such that well-known double connotations are possible which will gain of further importance, as shall be seen later.) Note that although E** is essentially acting on a different set of variables x*, we can say that E as well as E* (acting upon x) can be thought of as being *sublated* in E** in the threefold Hegelian sense: They are *annihilated* (because the process is acting now on a new level of evolution), but they are still *conserved* (because the new process has been constituted by the components of the old process), and they are thus *raised* to a higher level. On the other hand, the actual cycle implies that the procedure begins all over with a new set of amoebae which exist *on the same level* as the slime mold which is poducing them. Hence, although technically, the cycle is being characterized by a lowering of order again (as seen under the perspective of the new set of amoebae), *the complexity of the system as a whole has nevertheless been increased.* This will be also true for any of the following cycles. A triple {E, E*, E**} or alternatively, a double negation N$^2$, is called a *sandwich structure* and exhibits the threefold layer which is typical for an emergent process. (Note that this aspect has been described already in a qualitative language by Schelling, dating back as far as to the year 1831. [5])

## 1   Discrete Aspects of Negator Algebra

So far, the conception discussed above has centred on a continuous model of processes in the classical sense. The reason for this is that the discussion essentially emerged from earlier results of the Brussels school dealing with problems which would be appropriately modelled in terms of continuous equations of the reaction-diffusion type. However, the actual process cycle discussed above and represented by the negator action exhibits also a definitely discrete character, namely as a permanent succession of all these cycles which are „spiralling up-ward" through the levels of evolution (although we essentially deal with only to of them: of amoebae in general, and of slime molds). It is in fact the negation exponent which defines the *tact sequence* of the recursive action: because it is „counting" the numbers of self-compositions of the system which can be expressed as

$$dN(E)/ds = N^n(E).$$

This is nothing but a symbolic representation for the fact that the recursive sequence acts as self-composition satisfying the conditions of chaotic processes. (Cf. [1] again.) Hence, the term *ds* is *not* a differential, but indicates this formal succession of purely combinatorial steps which are being performed by successive realizations of the recursion. (This terminology however, secures the more than metaphorical analogy with dynamical systems.) The „spiral" of successive cycles (in the case of the slime mold e.g.) can be visualized then by a diagram of the form

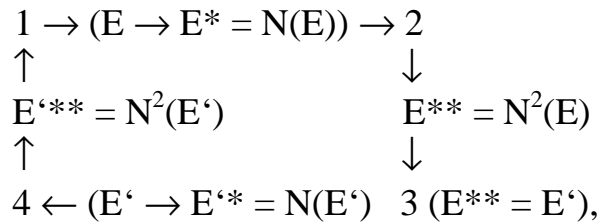

$$\begin{array}{lcl}
1 \to (E \to E^* = N(E)) & \to & 2 \\
\uparrow & & \downarrow \\
E'^{**} = N^2(E') & & E^{**} = N^2(E) \\
\uparrow & & \downarrow \\
4 \leftarrow (E' \to E'^* = N(E')) & & 3\ (E^{**} = E'),
\end{array}$$

meaning that transitions between states 1 through 3 form the first sandwich whose result is a stable structure on the new level of evolution. The transitions of the next sandwich (written with a prime here) lead back to the original situation, but they are „spiralling upward", because obviously, it is not the state 1 which is being attained following state 4, but some new state 1'. Hence, evolution shows up as a successive concatenation of sandwich structures.

These aspects are actually related to the concept of complexity. And this is already achieved in terms of the matrix representation we have chosen earlier: Remember that in the classical case (of continuous equations), the field of all possible equations can be written in abstract form as a network of interactions,

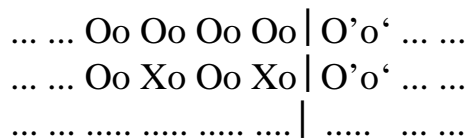

$$\begin{array}{l}
\ldots\ldots \text{Oo Oo Oo Oo} \mid \text{O'o'} \ldots\ldots \\
\ldots\ldots \text{Oo Xo Oo Xo} \mid \text{O'o'} \ldots\ldots \\
\ldots\ldots \ldots\ldots \ldots\ldots \ldots \mid \ldots\ldots \ldots\ldots
\end{array}$$

```
     ... ... ..... ..... ..... ....| ......  ... ...
     ... ... Oo Xo Oo Oo | O'o' ... ...
     ________________________________
     ... ... O'o' ... .. O'o'| O'o' ... ...,
```

where the capital letters indicate suitable coefficients and the small letters derivatives for equations of the general type

$$a(.)\, \partial^2/\partial x^2 + b(.)\, \partial/\partial x + c(.)\, x + ... + m(.)\, \partial^2/\partial y^2 + ... = 0.$$

In fact, there are two types of possible „innovation" due to the formation of structure: either there are „sleeping variables" whose coefficients have value zero in a given system such that their respective terms do not show up in the equations (this situation is indicated here by an X) – or, new derivatives are being spontaneously included in the system which have not been available before (this situation is indicated here by terms on the right-hand-side of the vertical lines or below the horizontal line). The „sleeping variables" represent what we might call an *internal potential* while the new variables represent an *external potential*. The idea is the following then: If actuality consists of everything that has been possible „before" (and is now actual), then it is nothing but a subset of the set of possible processes. But the question of whether the internal or the external potential is in fact actualized decides over the metaphysical issue of determinism: If only the former is being actualized, the „field of possibilities" itself is determined all the time. If the latter is being actualized, this field is not determined at all. It is the structure of the „matrix" which decides over this issue. I have discussed the far-reaching metaphorical implications of the concept of *matrix* at other places in some more detail. [6]
Note finally two more aspects: On the one hand, this matrix representation can serve to defining the concept of complexity in purely combinatorial terms: *Complexity* C shows up then as the quotient of actualized components and possible components, C := act/pos. Its complement is the *redundancy* R:= (pos – act)/pos. Hence, C + R = 1. In a physical system, we have maximal order (or minimal entropy), if C = 1, R = 0; and viceversa, we have minimal order (or maximal entropy), if C = 0, R = 1. Obviously, these definitions are compatible with what recent theories of self-organized criticality call *connectivity* of/in systems. [7] On the other hand, the matrix is nothing but a representation of a process simulation in terms of cellular automata, as can be easily seen when thinking of the „entrances" of coefficients-*cum*-variables as the sites of a cellular grid which can carry a certain „loading" with states and/or functions. [8] This is the reason that the slime mold cycle e.g. figures prominently in the StarLogo list of MIT's medialab. [9]

A final remark here: If comparing this conception with the notion of *algebraic action* introduced by Mike Manthey [10], then the parallel between one sequence of actions by boundary and co-boundary operators mediated in terms of what Manthey calls „twisted isomorphism" and one threefold sandwich layer of the negation cycle becomes straightforward. Because this idea of algebraic action aims to a description of *anticipatory systems* in the sense of artificial intelligence, what the twisted isomorphism actually mediates is the sensing of data (composition of structure from sensory input, increase of cognitive complexity) with the performing of practical actions (de-composition of structure, reduction of complexity). Hence, it is intrinsically *onto-epistemic*, meaning that it models actual processes (in ontological terms) and the actual modelling, the designing of these models (in epistemological terms), at the same time. This is what we know from category theory. Hence, the next section. Note also that in the recent terminology of self-organized criticality, Stuart Kauffman's „fourth law" of thermodynamics, stating that complexity increases always by actualizing those possibilities which are in the „adjacent possible" comprising of the set of all possible states which are one reaction step away from those which have been actualized before, can be easily translated into the language of the matrix here. As it appears, Kauffman would conjecture that only an internal potential should be available then. [11]

## 2  The Categorial Perspective

In the Keller-Segel example chosen above, the operators are acting on vectors which consist of particle number densities which in turn are functions of space and time, and of themselves. Visualized in terms of an appropriate category, we deal with objects which are states of a given system. We can call them *world states* provided we associate that given system with a „world" of its own. These states however are characterized by some suitable evolution operator which in our case shows up as a matrix. We call NEG the *negator category* whose objects are dynamical systems and whose morphisms are the negators themselves. Identities and compositions of morphisms are well-defined. Note however that there are no idempotents and well-defined inverses. (Note also that against common custom we have absorbed the time-derivative in the matrix representation of the actual operators.) Essentially, NEG has the structure of the category End of endomaps of sets whose objects are sets plus their endomaps and whose morphisms are structure preserving maps satisfying f o h = j o f, for any morphism f and endomaps h, j of two respective objects. Obviously, this is the suitable category describing sets of possible states of a system and their evolution by means of the endomaps and morphisms. For us, the morphisms are the negators. They tell us that if X and Y are any two objects in NEG, how to send a state x of X to a state which transforms under the dynamic of Y in the same way that x trans-forms under an endomap of X. The terminal object of this category describes the fixed points or equilibria. Note that in particular, the aforementioned „spiralling upward" of the sandwich cycle is itself a map in End.

Ideas of this sort have become relevant recently in quantum gravity, especially in a simplified variant of it called *topological quantum field theory* (TQFT). The basic point here is an important correspondence between the fundamental level of physics being associated with a purely combinatorial structure called *spin networks* originally invented by Roger Penrose, and generalized somewhat by Lee Smolin and Carlo Rovelli, and the „macroscopic" level of classical physics representing the phenomenological aspects of the world as they are being observed in practical terms of daily life. Details of this have been discussed at another place. [12] Here, we concentrate only on the correspondence procedure between the two levels of consideration. So what we essentially do is the following: Given the (n-1)-dimensional configuration space (of macroscopic phenomenology) S and a triangulation of S, choose a graph called the *dual 1-skeleton*. Express any state in *Fun* – which is the algebra of all functions on the space of connections of the principle G-bundle over space-time satisfying certain gauge conditions due to the gauge (Lie) group G – as a linear combination of states coming from spin networks whose underlying graph is this dual 1-skeleton. Define now space-time as a compact oriented cobordism M: S $\to$ S', where S, S' are compact oriented manifolds of dimension n-1. (Recall that two closed (n-1)-manifolds X and Y are said to be cobordant, if there is an n-manifold Z with boundary such that $\partial Z$ is the disjoint union of X and Y.) Choose a triangulation of M such that the triangulations of S, S' with dual 1-skeletons $\gamma$, $\gamma$' can be determined. The basis for gauge-invariant Hilbert spaces is given by the respective spin networks. Then the evolution operator Z(M): $L^2(A_\gamma/G_\gamma) \to L^2(A_{\gamma'}/G_{\gamma'})$ determines transition amplitudes <$\Psi$', Z(M) $\Psi$> with $\Psi$, $\Psi$' being spin network states. Write the amplitude as a sum over spin foams from $\Psi$ to $\Psi$': < , > = $\sum_{F:\Psi \to \Psi'}$ Z(F) plus composition rules such that Z(F') o Z(F) = Z(F' o F). This is a discrete version of a path integral. *Hence, re-arrangement of spin numbers on the „combinatorial level" is equivalent to an evolution of states in terms of Hilbert spaces in the „quantum picture" and effectively changes the topology of space on the „cobordism level"*. This can be understood as a kind of *manifold morphogenesis* in time: Visualize the n-dimensional manifold M (with $\partial$M = S $\cup$ S' - disjointly) as M: S $\to$ S', that is as a *process* (or as time) passing from S to S'. This the mentioned *cobordism*. Note that composition of cobordisms holds and is associative, but *not commutative*. The identity cobordism can be interpreted as describing a passage of time when topology stays constant. If there is no change of topology (due to the action of the identity cobordism), then there is no change of state, because we do not have any local degrees of freedom here. Visualized this way, TQFT might suggest that general relativity and quantum theory are not so different after all. In fact, *the concepts of space and state turn out to be two aspects of a unified whole, likewise space-time and process*. These results can also be formulated in the language of category theory: As TQFT maps each manifold S representing space to a Hilbert space Z(S) and each cobordism M: S → S' representing space-time to an operator Z(M): Z(S) → Z(S') such that composition and identities are preserved, this means that TQFT is a functor Z: nCob → Hilb. Note that the non-commutativity of operators in quantum theory corresponds to non-commutativity of composing cobordisms, and adjoint operation in quantum theory turning an operator A: H → H' into A*: H' → H corresponds to the operation of reversing the roles of past and future in a cobordism M: S → S' obtaining M*: S' → S.

But this is not simply a point of formal conceptualization. It is also an important aspect of the process of (epistemologically) *thinking about processes* itself. For the first time, Vladimir Trifonov has made this aspect explicit, as a consequence of applying topos-theoretic concepts to physics. [13] He introduces *topoi* (toposes)[3] as abstract worlds which represent universes of mathematical discourse whose inhabitants can utilize non-Boolean logics for their argumentation (i.e. their propositional structures): Be F a partially ordered field. Then, an F-*xenomorph* in the sense of Trifonov is a category A(F) of linear algebras over F. The objects of A(F) are called *paradigms* of an F-xenomorph, the arrows are called *actions*. Essentially, a paradigm then, is the set of states of knowledge. A paradigm is called *rational*, if the space of motions M(A) is a monoid. In particular, it can be shown that the set of all possible actions of a researcher is a topos whose arrows are those mappings which preserve realizations of the monoid (of the space of motions). It can also be shown that, if A is a rational paradigm, and the topos of all possible actions is Boolean (non-Boolean), then the paradigm A is classical (non-classical). For a xenomorph F = R of a generic type of the observer's psychology, Trifonov can also show that an R-xenomorph implies a classical Einstein paradigm, i.e. dimension 4 and signature 2 of the space-time metric. Also: If A is a non-trivial Grassmann algebra, then the paradigm is the Grassmannian of an R-xenomorph. Because A has a zero divisor, M(A) cannot be a group. Hence, the logic of a Grassmannian paradigm is always non-Boolean, and the mathematics is non-classical.

As I have discussed at another place [14], it is very likely that the category of negators NEG forms also a topos. It may even turn out that operator actions within NEG can be mapped by a functor into some suitable category such that this functor is basically identical with the functor Z: nCob → Hilb discussed above. The interesting point in the conception of Trifonov's is that (the logically formal part of) *thinking itself* is directly related to the physical process of unfolding the worldly structure as it can be described in terms of cosmological evolution. The basic idea in this is to define a self-referent cycle in the sense that the physical process is producing observers who choose their explicit logic for evaluating what they actually observe. We recognize the idea again, of nature exploring itself by means of human research (or telling its own story to itself, as a kind of self-narration which is modelling its own self-unfolding), which is an important aspect being recently discussed in philosophical terms. This is mainly due to the inherently logical structure of topoi. Take irreversibility for instance: Types of logic which have a modified law of negation, of the kind $\neg\,(\neg\,x) \neq x$, if x is a given proposition, determine the phenomenology as it is actually being observed according to the fact that processes are irreversible in the sense that their logical representation cannot reproduce initial propositions, independent of the number of negation operations acting repeatedly on such propositions. In other words: Recursive operations of this type have no fixed points globally, because they „spiral upward" through the levels of evolution. In a sense, we can say that temporality is coming in explicitly where earlier the logic remained static all the time (and created considerable difficulties when comparing theory with praxis, as e.g. Lacan has shown in some detail [15]). Hence, the advantage of topoi: They operate in terms of an intrinsic concept of time which can be visualized as a kind of generic concept, unifying the object level of a theory (that about which the theory is actually speaking) with the subject level (which determines the logic of the observer who actually speaks - in terms of the theory). This is another indication as to the phenomenological necessity of time on a macroscopic level of worldly perception and reflexion, thus in epistemological terms, while in ontological terms, time as a concrete variable can be absent.

---

[3] I keep to the original plural of „topos" used by Saunders MacLane, Goldblatt, and others. Besides being more correct in linguistic terms (because although used in French for the first time, and eventually being thought of as an abbreviation, its connotation is in fact a Greek one - which was also intended, by the way - hence, the Greek plural), it is also implying a nice double meaning, because in philosophical terms it has the meaning of characteristic, fundamental concepts (or categories). In ancient Greek, the word „category" is actually originating from *legal* language: Categorize (kathgore‹n) means „to accuse", and the categories are the actual points according to which a person is being accused in front of a judge and which are read from a list of such points. (Practically, humans are accusing nature, because apparently, it is of another *mode of being* than they themselves are.) Hence, categories are „topoi" of conflict.

## 3   Holistic Conclusions

We would not really like to interfere with quantum gravity. However, the important point we would like to make here is that these recent results of research in theoretical physics exhibit a lesson we can learn when studying evolutionary processes of a very different kind: When doing so what we have to bear in mind is that there are essentially two levels of reflexion, the fundamental one and the phenomenological (empirically observable) one. And although the detailed problems on the latter level may be of a very specific type, there is nevertheless their common ground which is to be found on the fundamental level of description. And this is an abstract but physical level which can be described in terms of a suitable mathematical language which in turn exhibits certain universal aspects of consideration. Hence, our holistic conclusions, in the following sense: With a view to a forthcoming project which is described in another paper [16], we have to actually differ between these two levels quite explicitly. The original problem which is at issue, namely the very practical question of studying a political project in the Italian city of Bologna – to close the historical centre of the city for any vehicle traffic (a problem generating the usual polarity of opinions) – has to be referred back to the problem of its foundation. That is, once the problem can be based on its own fundamental level from which it should be *derivable in principle* then, there is no room anymore for an emotional or intuitive argument in favour or against the issue, because it can be referred back to the common ground which governs all types of evolutionary processes of which the evolution of a city is only one small aspect. (As one can clearly recognize, this epistemological argument is based on the assumption of worldly rationality and at the same time loaded with many ethical implications – and this is what the modern philosophical attitude towards the sciences is actually all about.) So the idea is to visualize a city as a massively parallel network of self-organized emergent computational processes. And the objective should be to derive a similiar correspondence between the fundamental and the phenomenological level of the processes involved as it is the case in quantum gravity, where the theory itself can be represented by an appropriate (commutative) functor diagram of the form

$$\begin{array}{ccc} \text{nCob} & \to & \text{Hilb} \\ \uparrow & & \uparrow \\ \text{SpinF} & \to & \text{Hilb} \end{array}$$

where SpinF is the category of spin foams (evolutions of spin networks) whose objects are spin networks and whose morphisms are the spin foams themselves. The mapping on the right-hand-side indicates the identity functor on Hilb, while the lift mapping on the left-hand-side is hypothetical. Consequently, what we would find when treating our specific example, is an equivalent of the above diagram with a „right-hand-side completion" of the type

$$\begin{array}{cc} \leftarrow & \text{NEG} \\ & \uparrow \\ \leftarrow & \text{SpinF*}, \end{array}$$

where SpinF* is some „negating" equivalent of SpinF.

However, this kind of approach will not save us from having to do the „hard work" of actually modelling the city of Bologna (e.g.) in detail with a variety of available techniques. But what we can assume from the beginning on then, is that there are fundamental aspects common to all such processes which might help us to overcome obstacles to further insight into their structure which otherwise would not have been possible due to the restricted perspective taken with a view to the problem in question.

## Acknowledgements

For illuminating discussions on this topic I thank Anna Soci, Giorgio Colacchio (Bologna), Wolfram Voelcker (Berlin), and the members of the INTAS/NIS co-operation group project (INTAS 00-298) chaired by the indefatigable Wolfgang Hofkirchner (TU Vienna). I especially thank Anna Soci for the very kind invitation to the University of Bologna.